\documentclass[11pt,twoside]{article}
\usepackage{asp2006}
\usepackage{psfig}
\usepackage{epsf}
\usepackage{graphics}
\usepackage{lscape}
\def\edcomment#1{\iffalse\marginpar{\raggedright\sl#1\/}\else\relax\fi}
\reversemarginpar

\begin{document}
\title{Helioseismology over the Solar Cycle}
\author{M. J. Thompson}
\affil{School of Mathematics \& Statistics, University of Sheffield, 
S3~7RH, U.K.}

\def\etal{{\it et al.}}
\begin{abstract}
Helioseismology has produced unprecedented measurements of the Sun's internal
structure and dynamics over the past 25~years. Much of this work has been
based on global helioseismology. Now local helioseismology too is showing
its 
great promise. This review summarizes very 
briefly the principal global results that may be relevant to an understanding
of the origins of solar magnetism. Recent results regarding the variation of
frequencies over the solar cycle and the temporal variations of subsurface
flows are briefly summarized. 
\end{abstract}

\section{Introduction}

Helioseismology is concerned with the study of the Sun's internal structure 
and dynamics using the properties of acoustic waves that propagate 
through the interior and cause observable motion of the photosphere and lower 
solar atmosphere. The principal properties used for this study are the 
frequencies of resonant global modes of the Sun set up by these acoustic 
waves. Since the Sun is to a good approximation spherically symmetric, the
horizontal spatial structure of the modes is described by spherical 
harmonics $Y_l^m(\theta,\phi)$, where $\theta$ is co-latitude and $\phi$
is longitude. The modes are then labelled by three quantum numbers, the 
degree $l$ and order $m$ of the spherical harmonic and a radial order $n$ 
which is essentially the number of nodes in the mode's structure in the 
radial direction. The frequencies $\nu_{nlm}$ depend on the conditions in the
solar interior that affect wave propagation. In a non-rotating, perfectly 
spherically star the frequencies would have a degeneracy in that they would
not depend on $m$ for given $n$ and $l$: this degeneracy is lifted 
by rotation, structural asphericities and magnetic fields, and     
measurements of the resulting frequency splitting can be used to make 
inferences about these properties. 
(The frequency splitting 
within a multiplet of given $n$ and $l$ can be decomposed into
parts that are odd and even functions of $m$: the odd component arises from
rotation, while the even component arises from magnetic and 
thermal asphericities and distortions of the shape of the star from 
spherical symmetry.)
Application of inverse techniques provides
maps such as of the adiabatic sound speed $c$, density $\rho$, rotation and 
wave-speed asphericities in the Sun's otherwise impenetrable interior.

Spatially resolved measurements of the Sun's oscillations by the Global
Oscillation Network Group (GONG) and the Michelson Doppler Image (MDI) 
instrument on board the SOHO satellite began in the mid-1990s and thus now
provide essentially continuous coverage of one solar cycle. The 
whole-disk Sun-as-a-star measurements of the Birmingham Solar Oscillation 
Network (BiSON) extend back even further. Thus helioseismology is able to 
comment on frequency changes occurring over the solar cycle and possible
changes in flows and acoustic asphericities over that time.

\section{The mean solar structure and rotation}
Many of the results from helioseismology have been well described in 
reviews such as that by \cite{Christ2002}. Here we summarize briefly just
a few of the mean properties revealed by helioseismology that are pertinent to 
an understanding of the Sun's magnetic dynamo and the resulting activity 
cycle.

The Sun's convective envelope is very nearly adiabatically stratified, whereas
the radiative interior is subadiabatic. Thus the 
variation of the adiabatic sound speed with depth reveals where the
transition between the two occurs. Such analysis locates the base of the
convection zone at radius $r = 0.713\pm 0.003\,R_\odot$ from the centre of 
the Sun, where
$R_\odot$ is the Sun's radius \citep{Christ1991}. Note that this measures 
the extent of the essentially adiabatically stratified region, which may 
include a region of convective overshoot if the motions are sufficient to 
make that region adiabatically stratified. Simple models incorporating 
such overshooting typically have a rather sharp transition from 
adiabatic stratification to subadiabatic stratification in the radiative
interior: however, in the Sun the transition seems to be smoother than in 
the models \citep[see][]{Christ2010}.

Figure~1 shows the internal solar rotation over much of the convection zone
and outer radiative interior inferred from MDI data. The Sun rotates 
differentially throughout the convection zone, with a transition to what 
appears to be near-solid body rotation in the radiative interior. Between the
two regimes there is a rotational shear layer, the tachocline, which is now
widely considered to play a role in the large-scale solar dynamo. There is also
a near-surface shear layer which may also have a role. The tachocline is in 
fact narrower than the figure may suggest, because of the limited resolution 
of the inversion. Following first detailed quantification of the location
and extent of the tachocline by \cite{Kosov1996}, subsequent investigators have
mostly used a particular parametrization of the profile of the tachocline, 
giving its location and width \citep[e.g.][]{Charb1999} as
$0.693\pm 0.002 R_\odot$ and $0.039\pm 0.013 R_\odot$ at the equator: see these papers for
the precise meaning of these parameters. The tachocline is prolate, with the 
location different by about $0.02-0.03 R_\odot$ at $60^\circ$ latitude. Thus 
the bulk of the tachocline is in a stable subadiabatic region at low latitudes,
and straddles the base of the convection zone at higher latitudes.

 \begin{figure}[!h]
 \plotfiddle{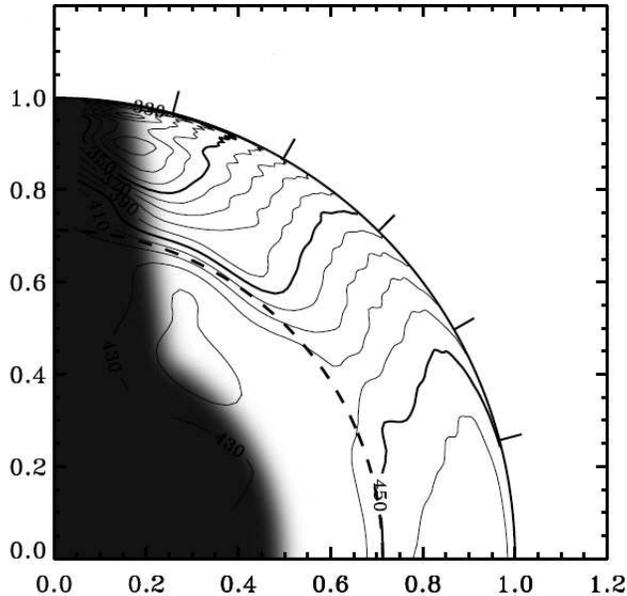}{3.4in}{0.}{40}{40}{-120}{0}
 \caption{Sun's internal rotation inferred from MDI data using a 2dSOLA
inversion \citep[from][]{Schou1998}.
Contours of iso-rotation are shown, with spacing $10\,$nHz, the bold contours being at $450$, $400$ and $350\,$nHz. The equator lies along the horizontal axis, and tick marks are at $15^\circ$ spacings in latitude. The base of the convection zone is indicated by the dashed line.  }
 \end{figure}

\section{Frequency variations over the cycle}

Since sound waves traverse the interior of the Sun in a time of order one hour, the global frequencies are determined by essentially the instantaneous state of 
the solar interior. If the internal structure or dynamics change over the 
course of the solar cycle, these changes may be reflected in changes in the
global frequencies. Indeed the frequencies and frequency splittings are 
observed to change over the cycle. Figure~2 shows the observed frequency 
shift in low-degree modes from BiSON observations, since 1985: the frequencies
vary by about $1$ part in $10^4$, the frequencies being highest at solar 
maximum. Also shown is
the variation over the same period in the $10.7\,$cm radio flux which is 
one widely used measure of solar activity. Clearly the frequency shifts and the
solar activity levels are very well correlated. 

Similarly tight correlations
have been demonstrated between the shifts in intermediate-degree modes and 
the photospheric magnetic flux, with the latitudinal distribution of 
the magnetic flux also plausibly explaining the even component of the 
frequency splitting varies with time \citep{Antia2001}. 

In conclusion, most if not all measured 
temporal variations in the mean frequencies and in the even component of the
frequency splittings are likely caused by surface changes in the magnetic
field, or by something that is highly correlated with them. There is little 
evidence for any contribution to the frequency changes over the solar
cycle
from structural or magnetic variations in the deeper interior.
However, the results in Fig.~2 hint that there may be something else 
going on: as reported by 
\cite{Broom2009},
there is 
some indication of a biennial oscillatory signal in the frequencies at all
times that is however only apparent in the $10.7\,$cm flux signal at high 
activity levels; compared with previous minima, the freqencies are even lower
at the present time relative to the $10.7\,$cm flux activity, 
and the correlation between frequencies and activity is less
good in the declining phase of the most recent cycle than at other times. All 
these aspects may indicate some contribution from the subsurface layers. There 
is also evidence of a change in the wave speed at the base of the convection
zone of about one part in $10^4$ between solar minimum and solar maximum
according to the analysis by \cite{Baldne2008}.

 \begin{figure}[!t]
 \plotfiddle{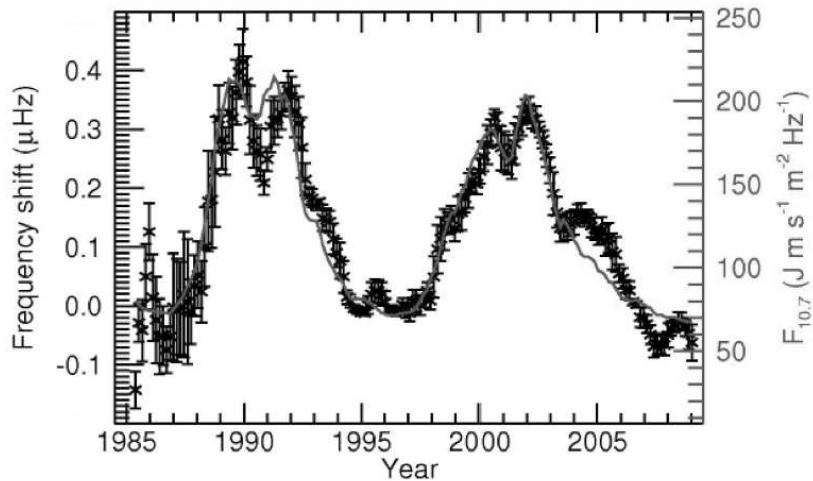}{2.6in}{0.}{40}{40}{-160}{0}
 \caption{Variation of mean low-degree frequencies over two solar cycles
\citep[from][]{Broom2009}.
Symbols with error bars show frequency shifts; the continuous curve (and the
right-hand scale) show corresponding levels of the $10.7\,$cm radio flux
over the same epoch. }
 \end{figure}

\section{Flow variations over the cycle}

The observed changes
in the odd component of the frequency splittings provide strong evidence 
for temporal variations in the subsurface rotation of the Sun. 
The so-called torsional oscillations -- weak but coherent banded zonal flows 
superimposed on the overall rotation profile -- were discovered in surface
Doppler measurements about three decades ago and have been shown by 
helioseismology to persist through a substantial fraction of the convection
zone \citep{Howe2000a}. Indeed, much of the convection zone seems to be 
exhibit angular velocity variations \citep{Voront2002}. In spite of the
present extended solar minimum, the equatorward 
migration of new prograde banded flows from mid-latitude is already well
underway, though its rate of migration is slower than it was during the 
previous minimum \citep{Howe2009}. Based on an analysis of these flows,
\citeauthor{Howe2009} estimate a length of approximately 12~years for Cycle~23.

A similar analysis to that of \cite{Howe2009} but extended to high latitudes
reveals further interesting behaviour of the rotation rate over time
(Howe {\it et al.} 2010, in preparation).  For example, 
at $75^\circ$ latitude the rotation rate has varied by almost $25\,$nHz over 
the cycle, reaching a minimum at around the start of 1999 and a maximum 
in 2003/4: for the past two years it pretty well repeated its behaviour of 11.5 years ago. At even higher latitudes the behaviour is rather similar, though
there is a curious double peak to the maximum, with one maximum in 2003 and 
the second one around the start of 2006: bearing in mind that these global
results do not separate out the northern and southern hemisphere, it is 
unclear at present whether the double-peak represents episodic behaviour or 
a difference in timing between the two hemispheres in reaching their maximum 
rotational speed. 

There 
is also evidence of a change in rotation rate near and possibly also beneath
the base of the convection
zone in the rising phase of the last cycle, with a period of about 1.3~years,
according to the analysis by \cite{Howe2000b}.

\section{Local helioseismology -- potential and issues}

Inversions based solely upon 
global mode frequencies have no longitudinal resolution,
nor can they distinguish the northern and southern hemispheres. Various local
helioseismic techniques offer the capability to study local features and 
different structures and flows in the two hemispheres. One such technique is
ring-diagram analysis (or simply ring analysis). Another is time-distance 
helioseismology. With these techniques it has proved possible to study 
the subsurface meridional flow and its variation over the solar cycle, but
thus far only in the outer few per cent by radius of the solar interior. 
Down to about $15\,$Mm the meridional flow is fairly uniform, of order 
$20\,$m/s, and poleward in direction, though with evidence of subsurface
counter-cells being present in the northern hemisphere around the time of solar
maximum and in the southern hemisphere during the declining phase of Cycle~23
\citep{Haber2006}. A measurement of the meridional circulation much deeper 
in the 
convection zone would be a valuable constraint on flux-transport dynamo
models \citep[e.g.][]{Dikpat2009}.
However, such a measurement is challenging: an estimate
indicates that the meridional flow at the base of the convection 
zone may not be detectable with measurements spanning an interval less than 
a solar cycle \citep{Braun2008}.

There is also strong evidence from local helioseismology
of signatures of evolving structures (thermal and/or 
magnetic) under active regions \citep[e.g.][]{Kosov2000}. However, there 
do appear to be significant discrepancies between the inferences obtained 
using  different methods \citep{Gizon2009} which indicates that a better
understanding of the forward modelling of the interaction of waves with 
magnetic structures and the resulting observables is required. 

\section{Conclusions}
Helioseismology has produced unprecedented measurements of the Sun's internal
structure and dynamics over the past 25~years. These include having mapped
the solar rotation over most of the interior, and discovering 
the solar tachocline.

The frequencies of the Sun's global oscillations change over the solar
cycle. Observed changes the odd component of the frequency splittings reflect 
changes in the solar internal rotation in possibly below the convection zone.
The behaviour of the banded zonal flows (torsional oscillations) give a 
length of approximately 12~years for Cycle 23.

Most of the changes in mean multiplet frequencies and in the even component 
of the frequency splittings comes from changes at or very close to the surface,
caused by changes in the surface magnetic field (or something that is 
highly correlated with the surface field). The frequencies of low-degree
modes are lower this minimum than during the previous minimum: there may be
some subsurface differences in the two minima that account for this. There
may also be some small variation in wave speed at the base of the convection
zone correlated with surface activity. 

Local helioseismology clearly detects temporal and spatial variations, and is
the fastest developing area of helioseismology. But there is a need for 
improved forward models in order to make robust inferences about the 
physical causes of those variations.

\acknowledgements
I thank Bernhard Fleck for financial support which made possible my participation at this
most stimulating meeting.  I thank Rachel Howe and Anne-Marie 
Broomhall in particular for sharing results with me for my presentation on
which this paper is based.


\begin{thebibliography}{}
\bibitem[Antia {{\etal}\/}(2001)]{Antia2001}
Antia, H. M., {\etal} 2001,
{\rm MNRAS}, {\rm 327}, 1029
\bibitem[Baldner \& Basu(2008)]{Baldne2008}
Baldner, C. S., \& Basu, S. 2008,
ApJ, 686, 1349
\bibitem[Braun \& Birch(2008)]{Braun2008}
Braun, D. C., \& Birch, A. C. 2008,
ApJ, 689, L161
\bibitem[Broomhall {{\etal}\/}(2009)]{Broom2009}
Broomhall, A. M., {\etal} 2009,
{\rm ApJ}, {\rm 700}, L162
\bibitem[Charbonneau {{\etal}\/}(1999)]{Charb1999}
Charbonneau, P., {\etal} 1999,
{\rm ApJ}, {\rm 527}, 445
\bibitem[Christensen-Dalsgaard(2002)]{Christ2002}
Christensen-Dalsgaard, J. 2002,
{\rm Rev. Mod. Phys.}, {\rm 74}, 1073 
\bibitem[Christensen-Dalsgaard {{\etal}\/}(1991)]{Christ1991}
Christensen-Dalsgaard, J., Gough, D. O., \& Thompson, M. J. 1991,
{\rm ApJ}, {\rm 378}, 413 
\bibitem[Christensen-Dalsgaard {{\etal}\/}(2010)]{Christ2010}
Christensen-Dalsgaard, J., Monteiro, M. J. P. F. G., Rempel, R., \& 
Thompson, M. J. 2010,
{\rm MNRAS}, to be submitted
\bibitem[Dikpati \& Gilman(2009)]{Dikpat2009}
Dikpati, M., \& Gilman, P. A. 2009,
{\rm Sp. Sci. Rev.}, {\rm 144}, 67
\bibitem[Gizon {{\etal}\/}(2009)]{Gizon2009}
Gizon, L., {\etal} 2009,
{\rm Sp. Sci. Rev.}, {\rm 144}, 249
\bibitem[Haber {{\etal}\/}(2006)]{Haber2006}
Haber, D., Hindman, B., Toomre, J., \& 
Bogart, R. S. 2006,
{\rm ESA SP-624}, CDROM~p.45.1
\bibitem[Howe {{\etal}\/}(2000a)]{Howe2000a}
Howe, R., {\etal} 2000a,
{\rm ApJ}, {\rm 533}, L163
\bibitem[Howe {{\etal}\/}(2000b)]{Howe2000b}
Howe, R., {\etal} 2000b,
{\rm Science}, {\rm 287}, 2456 
\bibitem[Howe {{\etal}\/}(2009)]{Howe2009}
Howe, R., {\etal} 2009,
{\rm ApJ}, {\rm 701}, L87 
\bibitem[Kosovichev(1996)]{Kosov1996}
Kosovichev, A. G. 1996,
{\rm ApJ}, {\rm 469}, L61
\bibitem[Kosovichev {{\etal}\/}(2000)]{Kosov2000}
Kosovichev, A. G., Duvall, T. L., Jr., \& Scherrer, P. J. 2000,
{\rm Sol. Phys.}, {\rm 192}, 159
\bibitem[Schou {{\etal}\/}(1998)]{Schou1998}
Schou, J., {\etal} 1998,
{\rm ApJ}, {\rm 505}, 390 
\bibitem[Vorontsov {{\etal}\/}(2002)]{Voront2002}
Vorontsov, S. V., {\etal} 2002,
{\rm Science}, {\rm 296}, 101 

\end{thebibliography}
\end{document}